\documentclass{article}
\usepackage{graphicx}
\usepackage{citesupernumber}
\usepackage{dcolumn}
\usepackage{bm}
\usepackage{amsmath}
\usepackage{amssymb}
\usepackage[rflt]{floatflt}
\usepackage{latexsym}
\usepackage[dvips]{color}
\usepackage{psfrag}
\usepackage{natbib}
\usepackage{times}
\makeatletter

\def\frtnsfb{\fontfamily{\sfdefault}\fontseries{bx}\fontshape{n}\fontsize{14.4}{18pt}\selectfont}
\def\svtnsfb{\fontfamily{\sfdefault}\fontseries{bx}\fontshape{n}\fontsize{17.28}{22pt}\selectfont}
\def\twtysfb{\fontfamily{\sfdefault}\fontseries{bx}\fontshape{n}\fontsize{20.74}{25pt}\selectfont}
\def\twfvsfb{\fontfamily{\sfdefault}\fontseries{bx}\fontshape{n}\fontsize{24.88}{30pt}\selectfont}
\def\large{\@setsize\large{18pt}\xivpt\@xivpt \let\sfb=\frtnsfb}
\def\Large{\@setsize\Large{22pt}\xviipt\@xviipt \let\sfb=\svtnsfb}
\def\LARGE{\@setsize\LARGE{25pt}\xxpt\@xxpt \let\sfb=\twtysfb}
\def\huge{\@setsize\huge{30pt}\xxvpt\@xxvpt \let\sfb=\twfvsfb}
\makeatother

\begin{document}
\parindent=10mm
 \baselineskip=8mm
 \noindent {\Large\sfb On the effect of fluctuating recombination rates
 on the decorrelation of gene histories in the human genome }\\[1cm]
 A. Eriksson and B. Mehlig\\[1cm]
 \mbox{}
 {\em Physics and Engineering Physics,  Gothenburg University/Chalmers,
 41296 Gothenburg, Sweden}
 \\[1cm]
 \begin{abstract}
  \baselineskip=8mm
  We show how to incorporate fluctuations of the recombination
  rate along the chromosome into  standard gene-genealogical
  models for the decorrelation of gene histories. This enables us to
  determine how
  small-scale fluctuations (Poissonian hot-spot model)
  and large-scale variations \citep{kon02} of the
  recombination rate influence 
  this decorrelation.
  We find that the empirically determined large-scale variations
  of the recombination rate give rise to a significantly slower 
  decay of correlations compared to the standard, unstructured
  gene-genealogical
  model assuming constant recombination rate. A model with long-range
  recombination-rate variations and with demographic
  structure (divergent population) 
  is found to be consistent with the empirically observed
  slow decorrelation of gene histories. Conversely,
  we show that small-scale recombination-rate
  fluctuations do not alter the large-scale decorrelation of
  gene histories.
  \end{abstract}

\mbox{}\\
  Genome-wide variation and decorrelation of gene histories are reflected in
  patterns of linkage disequilibrium which in turn shape the
  genetic variation observed on the molecular level. Recently
  \citet{rei02} reported on the first genome-wide measurement of
  correlations of human gene histories. \citet{rei02} show that
  their data are inconsistent with standard gene-genealogical models
  allowing for non-trivial population structures and demographic
  schemes, but assuming a constant recombination rate over the
  genome. 
  The question is thus: can fluctuations of the
  recombination rate along the chromosome  explain the slow
  correlation decay of gene histories? 
  
  Empirical results \citep{cha84,gol01,jef01} indicate that
  the recombination rate is not constant along the chromosome. It
  was observed that, at certain locations,  an appreciable fraction
  of recombination events are concentrated in short regions [roughly
  $1$kb long and spaced $50$kb apart \citep{stu03}], so-called hot
  spots. At least locally this implies small-scale ($<100$kb)
  variations of the recombination rate along the chromosome. 
  Genome-wide, long-range fluctuations of the recombination rate for
  humans have been empirically determined by \cite{kon02}.
  It is thus necessary to incorporate the effect of fluctuating
  recombination rates into the standard gene-genealogical
  model \citep{gri81,hud83,tav84,kap85,hud90,nor02}. More generally,
  it is necessary to determine: on which length scales
  do recombination-rate fluctuations at a certain scale 
  influence the decorrelation function of gene histories most
  significantly?
  It has been argued \citep{rei02} that small-scale recombination-rate
  fluctuations ($< 100$kb) related to hot spots
  are an important if not the main feature determining the slow
  decorrelation of gene histories (assuming
  that hot-spots are to be found genome-wide).

  Here we derive an expression for the correlation of gene histories
  in neutral gene-genealogical models
  allowing for fluctuating recombination rates.
  This enables us to explain and quantitatively describe the
  influence of recombination-rate fluctuations
  on the correlation of gene histories.
  We find that large-scale fluctuations of empirically
  determined recombination rates \citep{kon02} 
  give rise to a significantly slower 
  decay of correlations compared to the standard, unstructured,
  constant population-size
  gene-genealogical model assuming constant recombination rate.
  Furthermore, a model with large-scale recombination-rate fluctuations
  and with demographic structure 
  [divergent population, see \citep{maize,tes02,rei02} and references cited
  therein] 
  is found to be consistent with the
  empirically observed decorrelation of gene histories.
  It is not necessary to invoke hot
  spots.

  In a neutral model,  \citet{kap85} [see also \citep{gri81}] have
  derived
  a relation between the correlation function $\rho_{\tau_x,\tau_y}$ of
  the times $\tau_x$ and $\tau_y$ to the most
  recent common ancestors of two loci $x$ and $y$,
  and the amount $C$ of recombination between these two
  loci\footnote{The
  result of  \citet{kap85} for the  unstructured,
  constant
  population-size model is exact for sample size $n=2$; for large $n$
  it is
  a very good approximation.}.
  We observe that their result depends on the total amount of
  recombination
  between the two loci, but not on the distribution of recombination
  events between these loci. Moreover, this is still true
  when population structure is taken into account.  
  The expected correlation  $\rho^{\rm exp}_{\tau_x,\tau_y}$
  is obtained by averaging with a sliding window of
  length $|y-x|$ along the chromosome.
  Thus, if $p_X(C)$ is the genome-wide distribution
  of recombination intensity $C$
  in bins of lengths $X=|y-x|$, the expected correlation is
  \begin{equation}
  \label{eq:2}
  \rho^{\rm exp}_{\tau_x,\tau_y} = \int\!{\rm d}C\, p_X(C)\,
  \rho_{\tau_x,\tau_y}(C)\,.
  \end{equation}
  It also follows that
  small-scale
  fluctuations of the recombination rate on length scales much smaller
  than $X$ are irrelevant to the decay of correlations on scales of the
  order of $X$. In particular, fluctuations due to hot spots at small scales
  cannot change the decorrelation of gene histories
  at much larger scales.

\begin{figure}[b]
  \begin{minipage}{5cm}
  \psfrag{a}{\Large\sfb a }
  \psfrag{R1}{$R_1$}
  \psfrag{R0}{$R_0$}
  \psfrag{l}{\hspace*{-0.2cm}$\lambda^{-1}$}
  \psfrag{y}{\hspace*{-1cm}$R(x)$}
  \psfrag{x}{$x$}
  \hspace*{1.05cm}\includegraphics[width=6.5cm,clip]{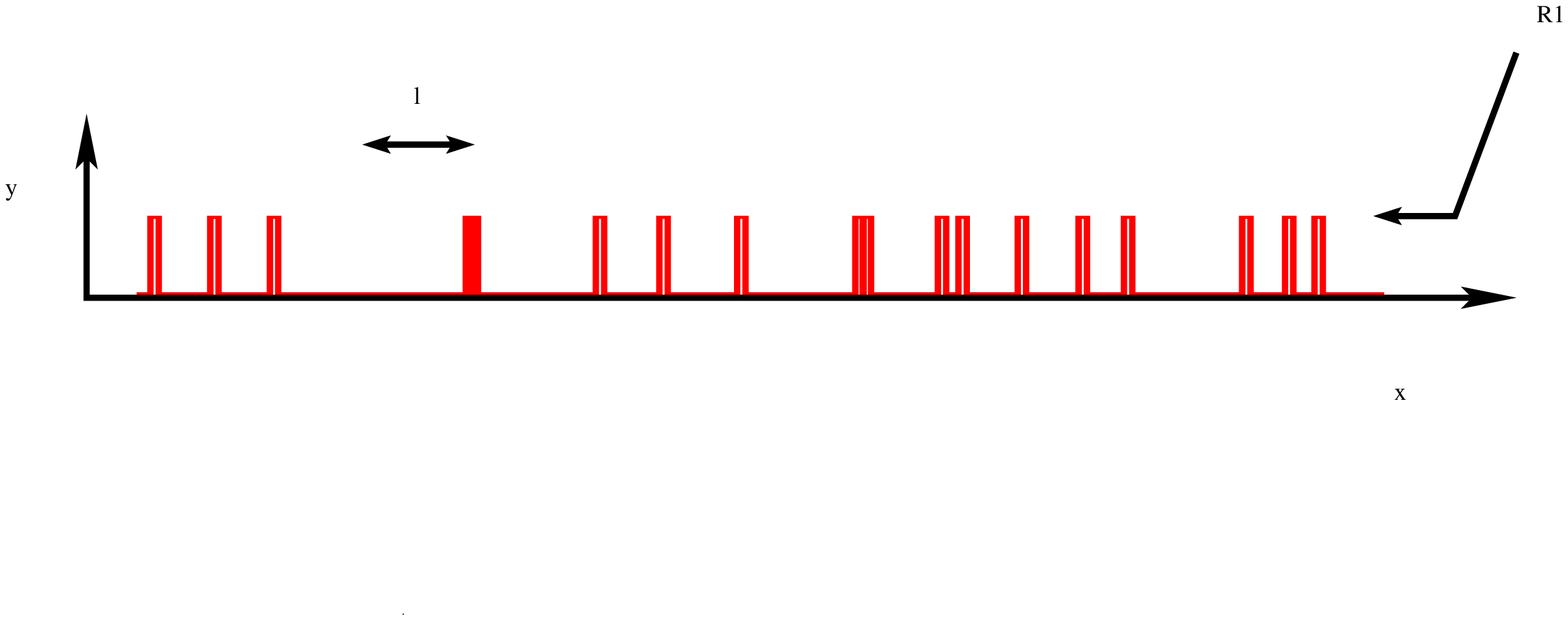}
  \mbox{} \\[1.5cm]
  \psfrag{b}{\Large\sfb b }
  \psfrag{y}{\hspace*{-7mm}$R(x)$}
  \psfrag{x}{\raisebox{-2mm}{$x$}}
  \psfrag{t1}{\small hot-spot cluster, length $pL$}
  \psfrag{t2}{\small region of constant $R$, length}
  \psfrag{t3}{\hspace*{2.25mm}\raisebox{-2mm}{\small $(1\!-\!p)L$}}
  \hspace*{1.4cm}\raisebox{3mm}{\includegraphics[width=5cm,clip]{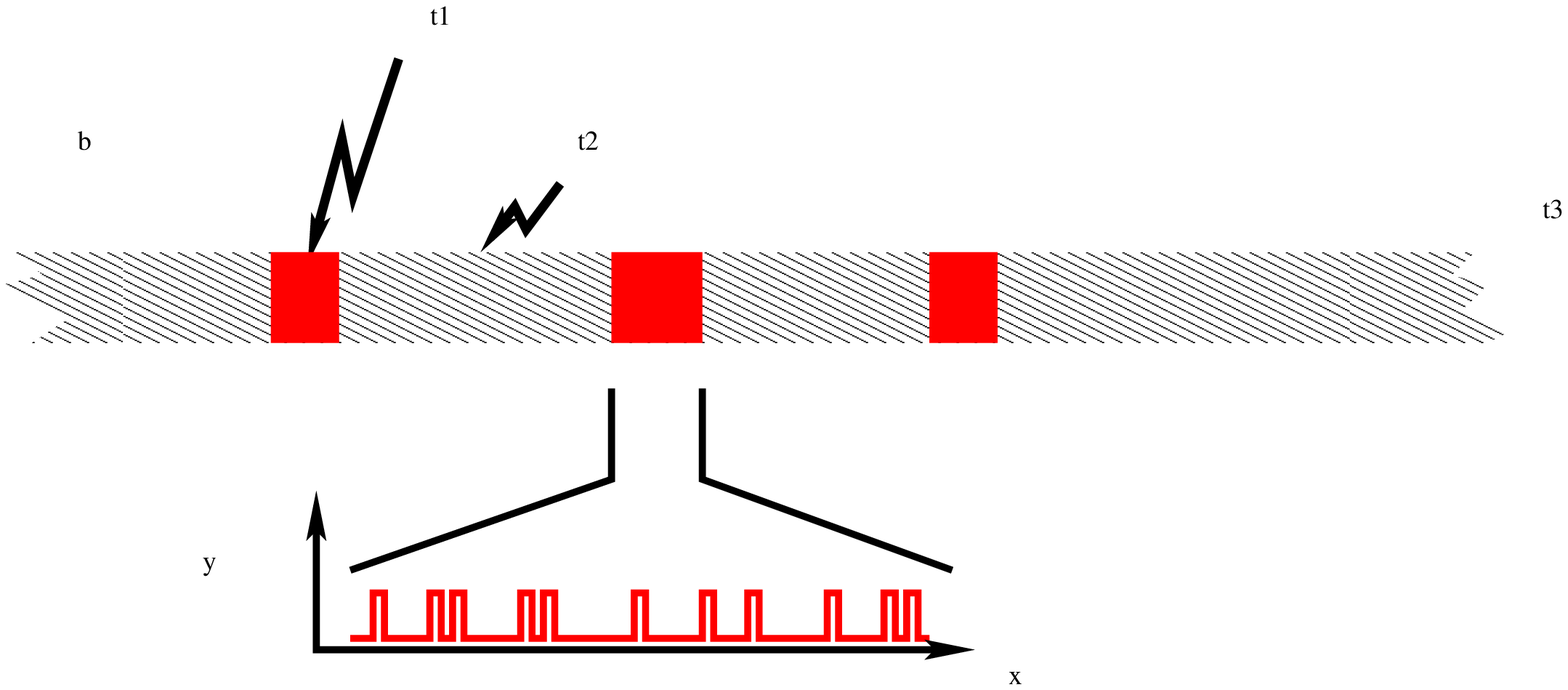}}\end{minipage}
   \begin{minipage}{5cm}
  \psfrag{d}{\hspace*{1.1cm}\raisebox{-8mm}{\Large\sfb c }}
  \psfrag{t2}{$10^3$}
   \psfrag{x}{\hspace*{.2cm} $C$}
   \psfrag{y}[c][Bl][1][90]{\mbox{}\hspace*{10mm}\raisebox{-2mm}{$p_X(C)$}}
   \psfrag{t1}[c][Bl][1][90]{\hspace*{5mm}$10^{-2}$}
\psfrag{y0}[c][Bl][1][90]{$0$}
\psfrag{x0}{$0$}
  \hspace*{4.5cm}\raisebox{-4mm}{\includegraphics[width=3.5cm,clip]{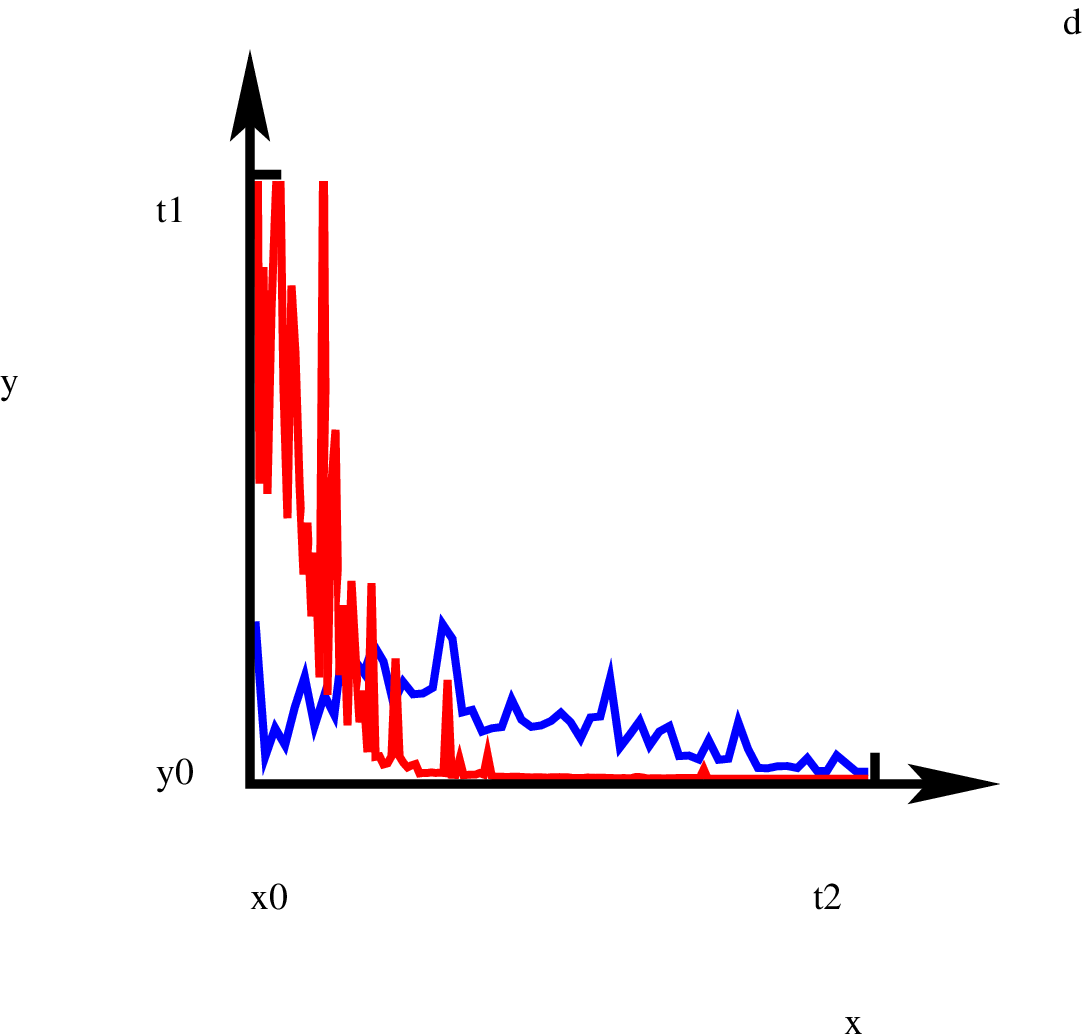}}
  \mbox{}\\[2cm]
  \psfrag{c}{\raisebox{8mm}{\Large\sfb d }}
   \psfrag{t5}[c][Bl][1][90]{time $\tau$}
   \psfrag{t1}{$N$}
   \psfrag{t2}{$\!\!\!\gamma N$}
   \psfrag{t3}{$\!(1\!-\!\gamma) N$}
   \psfrag{t4}{$\tau_0$}
  \hspace*{4.5cm}\raisebox{-4mm}{\includegraphics[width=5cm,clip]{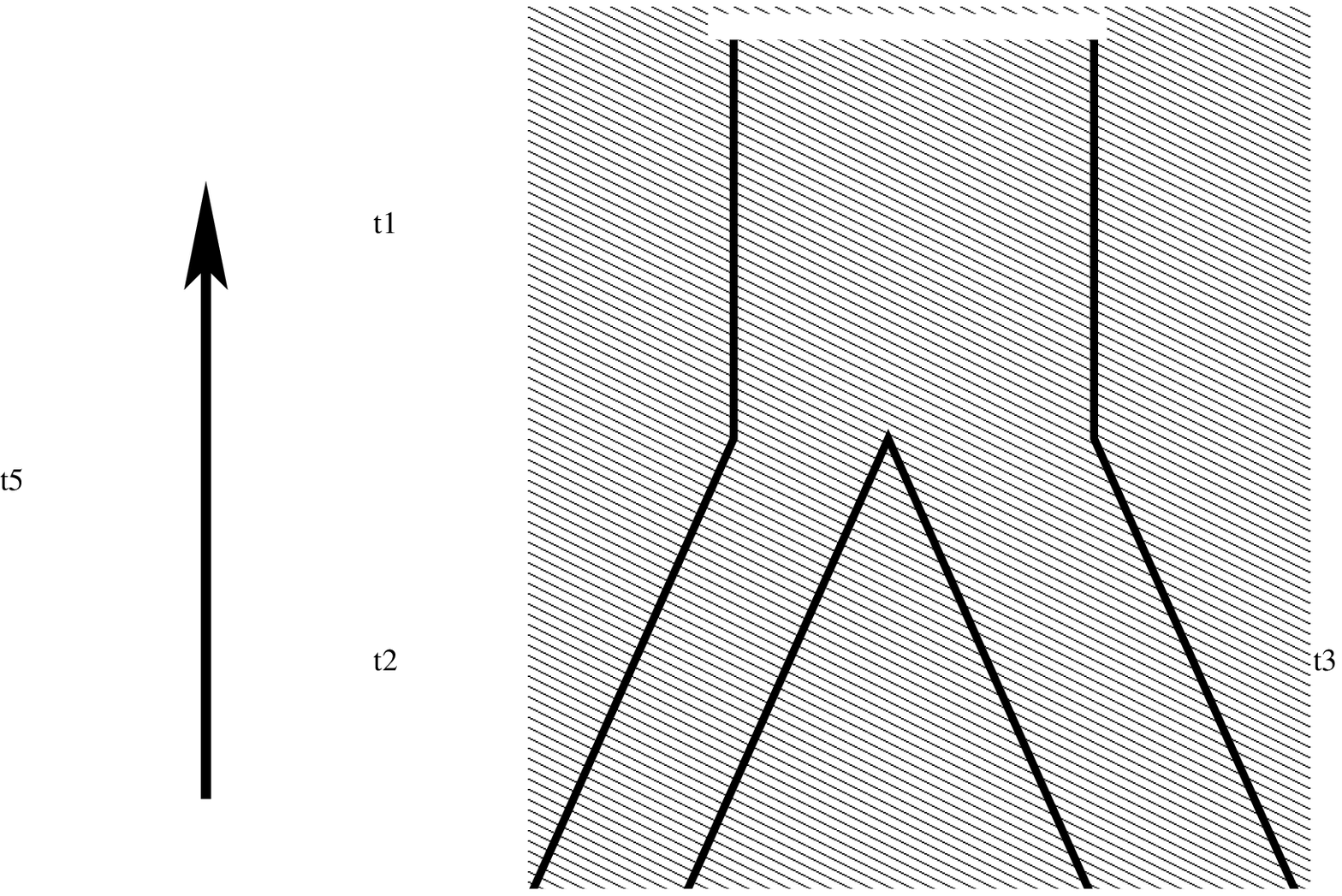}}\end{minipage}
  \mbox{}\\[1cm]
  \caption{\label{fig:1}
   \baselineskip=8mm
    Description of models.
    {\bf a} Model I:
    Recombination events occur at hot-spots (of zero width) with rate
    $R_1=R/\lambda$. 
    The remaining fraction
     occurs uniformly with rate $R_0=(1\!-\!p)R$.
     The number of hot-spots in a locus of length $l$ is Poisson
     distributed with rate $\lambda l$.
    Eq. (\ref{eq:2}) gives
    $ \rho^{\rm exp}_{\tau_x,\tau_y}\!=\!
    {\rm e}^{-\lambda X} \sum_{k=0}^\infty {(\lambda X)^k}/{k!}\,%
    (R_1 k\!+\!18)%
    /[(R_1 k)^2\!+\!13\,R_1 k\!+\!18] $. This result is exact for $n=2$.
    {\bf b} Model II: hot-spots occur in clusters of size $pL$,
    separated by empty regions of length $(1\!-\!p)L$, $0\leq p \leq
    1$, and $L$ is a typical length scale (of the order of several
    Mb). Within a cluster, the number of hot-spots is Poisson distributed.
    {\bf c} Model III: the genome-wide distribution 
    $p_X(C)$ is obtained by sampling $C=g(x+X)-g(x)$
    from empirical data \citep{kon02}
    on the cumulative genetic distance
    $g(x)$ by randomly choosing physical positions $x$.
    The curve $g(x)$ is obtained from the Nature Genetics web supplement
    NG917-S13 \citep{kon02}, from columns 1 (physical distance) and 3 
    (sex-averaged genetic distance) assuming an effective population
    size
    of $N=10^4$, by ignoring entries labeled ``NA",
    and by shifting
    the origin of both physical and genetic distances so that $g(0)=0$.
    Shown here is $p_X(C)$ for chromosome 5; $X=200$kb (red) and $1$Mb 
    (blue).
    {\bf d} Demographic structure (divergent population):
    it is assumed that the population
    size was $N$ until time $\tau_0$ in the past when it split into
    two populations of sizes $\gamma N$ and $(1-\gamma)N$. 
    The parameters $\tau_0$ and $N$ are chosen to be
    consistent with empirical data on the mean of the time
    to the most common recent ancestor and its coefficient
    of variation; table 1 in \citep{rei02}. The asymmetry parameter
    varies between $0$ and $1/2$. For sample size $n=2$, the function
    $\rho_{\tau_x,\tau_y}(C)$ for this model was calculated by
    \cite{eri03}.
    }
    \end{figure}

  Using (\ref{eq:2})
  we have computed
  $\rho^{\rm exp}_{\tau_x,\tau_y}$ in four models (Fig.~\ref{fig:1}):
  assuming small-scale variation of the recombination rate (model I),
  incorporating, in addition,
  large-scale variation (model II),
  and estimating $p_X(C)$ from the empirical data of \citet{kon02}
  (model III), and, in addition, taking into account demographic
  population structure (model IV).
  Model I is the Poissonian hot-spot model of \citet{rei02},
  described in more detail in Fig.~1a below. From (\ref{eq:2}) we obtain
  an explicit
  expression for $\rho^{\rm exp}_{\tau_x,\tau_y}$ (caption of
  Fig.~1a). This result is
  shown in Fig.~2a.
  In agreement with \citet{rei02}
  we find that on distances of the order of the hot-spot spacing,
  correlations are larger than those in a constant recombination-rate
  model.
   However, there is no choice of parameters which could explain the
   empirically observed
   decorrelation function.
   (c.f. data in  Fig.~6a of \citet{rei02}, reproduced in Fig.~2b
   below). In particular, no significant increase in correlations
   on length scales $\gg \lambda^{-1}$ is observed, as discussed above.

  \citet{rei02} have fitted an ``arbitrary mixed model" to their
  empirical
  data. In order to obtain these results it is necessary to introduce
  large-scale variations of the recombination rate, on a scale $L\gg
  X\sim 1$Mb.
  One possibility (model II) is to assume that hot-spots occur in
  clusters, with long ($\gg 1$Mb)  regions of low recombination
  intensity
  between them, see Fig.~1b. This model provides a better fit to the
  empirical data (Fig.~2b)
  than model I, indicating that large-scale fluctuations
  of the recombination rate are important.
  Notice that assuming $p_X(C) = (1-p)\,\delta(C-R_0X) +
  p\,\delta(C-R_1X)$
  can produce an equally good fit to the data (e. g. for $p=0.55$, $R_0 =
  1.2$cM/Mb and $R_1=0.02$cM/Mb, not shown). In this model
  the recombination rate is constant on large scales ($\gg 1$Mb)
  and alternates between two values $R_0$ and $R_1$. However 
  this model is not consistent with the empirically observed $p_X(C)$.
  We have estimated $p_X(C)$ from empirical data
  \citep{kon02} (Fig.~\ref{fig:1}c, model III). 
  The corresponding results are shown in
  Fig.~2b. We find that the empirically determined large-scale
  fluctuations of the recombination rate give rise to
  significantly enhanced correlations (compared to the standard
  model assuming constant recombination rate), especially at
  large distances.

  It is expected that population
  structure can increase the correlations of gene histories at large
  distances.
  We have considered the effect of large-scale recombination-rate
  fluctuations within a well-established model of demographic structure:
  the population was of constant
  size $N$ until $\tau_0$ generations ago, when it split into two
  fractions of size $\gamma N$ and $(1 - \gamma)N$. The two
  sub-populations remained separate until a recent merging
  (see for instance \cite{maize,tes02,rei02} and references therein).
  For sample size $n=2$ we have calculated $\rho_{\tau_x,\tau_y}(C)$ explicitly
  in this model \citep{eri03}.
  Without recombination-rate fluctuations,
  this model does not describe the empirically observed
  correlation of gene histories \citep{rei02}.
  We have determined the effect of 
  large-scale recombination-rate fluctuations \citep{kon02} 
  on the correlation of gene histories in this model using eq. (1)
  and the explicit expression for $\rho_{\tau_x,\tau_y}(C)$.
  The parameters of the model ($\tau_0$ and $N$) where chosen to be
  consistent with the empirically estimated time to the most
  recent common ancestor and its coefficient of variation \citep{rei02}.
  The parameter $\gamma$ was set to $0.3$.
  The resulting correlation function
  matches the empirical data reasonably well. 
  Decreasing $\gamma$ gives rise to decreased correlations
  ($\gamma=0$ corresponds to the standard model without demographic
  structure).

  In summary we have determined the influence of recombination-rate
  fluctuations on the decorrelation of gene histories.
  We find that small-scale fluctuations are irrelevant to
  long-range correlation decay. Empirically determined
  large-scale fluctuations of the recombination rate, however,
  are found to significantly increase the correlations.
  Within a model with demographic structure,  large-scale
  fluctuations of empirically determined recombination rates
  significantly contribute to the empirically observed slow
  decay of correlations.

  We conclude by discussing the implications of our results
  for the study of genome-wide variability as reflected in
  single-nucleotide polymorphism (SNP) statistics.
  Eq. (1) determines the effect of recombination-rate fluctuations
  on $\rho_{\tau_x,\tau_y}^{\rm exp}$.  
  This quantity, in turn, determines the genome-wide statistics
  of SNP locations:
  the variance of the number of SNPs in bins of lengths $l$
  along the chromosomes is determined by the integral
  of $\rho_{\tau_x,\tau_y}^{\rm exp}$ over $x$ and $y$ from
  $0$ to  $l$,
  i.e. by how fast the correlations decay on scales of
  length $l$ \citep{hud90}. \citet{tsc01} has empirically determined
  the variance of the number of SNPs in short reads
  (of average length $500$bp), the result was found to be consistent
  with the standard, unstructured gene-genealogical model assuming a constant
  recombination rate. This is consistent with our results (Fig. 2b):
  on scales of the order of $500$bp, the recombination-rate fluctuations
  have little effect on the correlation function.
  We expect, however, that in order to understand the statistics 
  of SNP counts in longer bins, it will be necessary to account for
  long-range recombination-rate fluctuations.

  \mbox{}\\[-2cm]
  \bibliography{refs}

\begin{thebibliography}{16}
\expandafter\ifx\csname natexlab\endcsname\relax\def\natexlab#1{#1}\fi

\bibitem[{\sc Chakravarti} {\em et~al.\/}, 1984]{cha84}
{\sc Chakravarti, A.}, {\sc Buetow, K.~H.}, {\sc Antonarakis, S.~E.}, {\sc
  Waber, P.~G.}, {\sc Boehm, C.~D.}, and {\sc Kazazian, H.~H.}, 1984 Nonuniform
  recombination within the human $\beta$-globin gene cluster.
\newblock Am. J. Human Genetics {\bf 36}: 1239--1258.

\bibitem[{\sc Eriksson} and {\sc Mehlig}, 2004]{eri03}
{\sc Eriksson, A.} and {\sc Mehlig, B.}, 2004 unpublished .

\bibitem[{\sc Eyre-Walker} {\em et~al.\/}, 1998]{maize}
{\sc Eyre-Walker, A.}, {\sc Gaut, R.~L.}, {\sc Hilton, H.}, {\sc Feldman,
  D.~L.}, and {\sc Gaut, B.~S.}, 1998 Investigation of the bottleneck leading
  to the domestication of maize.
\newblock PNAS {\bf 95}: 4441--4446.

\bibitem[{\sc Goldstein}, 2001]{gol01}
{\sc Goldstein, D.~B.}, 2001 Islands of linkage disequilibrium.
\newblock Nature Genetics {\bf 29}: 109--111.

\bibitem[{\sc Griffiths}, 1981]{gri81}
{\sc Griffiths, R.~C.}, 1981 Neutral $2$-locus multiple allele models with
  recombination.
\newblock Theor. Pop. Biol. {\bf 19}: 169--186.

\bibitem[{\sc Hudson}, 1983]{hud83}
{\sc Hudson, R.~R.}, 1983 Properties of a neutral allele model with intragenic
  recombination.
\newblock Theor. Pop. Biol. {\bf 23}: 183--201.

\bibitem[{\sc Hudson}, 1990]{hud90}
{\sc Hudson, R.~R.}, 1990 Gene genealogies and the coalescent process.
\newblock Oxford Surveys in Evolutionary Biology {\bf 7}: 1--44.

\bibitem[{\sc Jeffreys} {\em et~al.\/}, 2001]{jef01}
{\sc Jeffreys, A.~J.}, {\sc Kauppi, L.}, and {\sc Neumann, R.}, 2001 Intensely
  punctate meiotic recombination in the class ii region of the major
  histocompatibility complex.
\newblock Nature Genetics {\bf 29}: 217--222.

\bibitem[{\sc Kaplan} and {\sc Hudson}, 1985]{kap85}
{\sc Kaplan, N.} and {\sc Hudson, R.~R.}, 1985 The use of sample genealogies
  for studying a selectively neutral $m$-loci model with recombination.
\newblock Theor. Pop. Biol. {\bf 28}: 382--396.

\bibitem[{\sc Kong} {\em et~al.\/}, 2002]{kon02}
{\sc Kong, A.}, {\sc Gudbjartsson, D.~F.}, {\sc Sainz, J.}, {\sc Jonsdottir,
  G.~M.}, {\sc Gudjonsson, S.~A.}, {\sc Richardsson, B.}, {\sc Sigurdardottir,
  S.}, {\sc Barnard, J.}, {\sc Hallbeck, B.}, {\sc Masson, G.}, {\sc Shlien,
  A.}, {\sc Palsson, S.~T.}, {\sc Frigge, M.~L.}, {\sc Thorgeirsson, T.~E.},
  {\sc Gulcher, J.~R.}, and {\sc Stefansson, K.}, 2002 A high-resolution
  recombination map of the human genome.
\newblock Nature Genetics {\bf 31}: 241--247.

\bibitem[{\sc Nordborg} and {\sc Tavar{\'e}}, 2002]{nor02}
{\sc Nordborg, M.} and {\sc Tavar{\'e}, S.}, 2002 Linkage disequlibrium: what
  history has to tell us.
\newblock TRENDS in Genetics {\bf 18}: 83--85.

\bibitem[{\sc Reich} {\em et~al.\/}, 2002]{rei02}
{\sc Reich, D.~E.}, {\sc Schaffner, S.~F.}, {\sc Daly, M.~J.}, {\sc McVean,
  G.}, {\sc Mullikin, J.~C.}, {\sc Higgins, J.~M.}, {\sc Richter, D.~J.}, {\sc
  Lander, E.~S.}, and {\sc Altshuler, D.}, 2002 Human genome sequence variation
  and the influence of gene history, mutation and recombination.
\newblock Nature Genetics {\bf 32}: 135--142.

\bibitem[{\sc Stumpf} and {\sc Goldstein}, 2003]{stu03}
{\sc Stumpf, M. P.~H.} and {\sc Goldstein, D.~L.}, 2003 Demography,
  recombination hotspot intensity, and the block structure of linkage
  disequilibrium.
\newblock Current Biology {\bf 13}: 1--8.

\bibitem[{\sc Tavar{\'e}}, 1984]{tav84}
{\sc Tavar{\'e}, S.}, 1984 Lines of descent and genealogical processes, and
  their application in population genetics models.
\newblock Theor. Pop. Biol. {\bf 26}: 119--164.

\bibitem[{\sc Teshima} and {\sc Tajima}, 2002]{tes02}
{\sc Teshima, K.} and {\sc Tajima, F.}, 2002 The effect of migration during the
  divergence.
\newblock Theor. Pop. Biol. {\bf 62}: 81--95.

\bibitem[{\sc {The International SNP Map Working Group}}, 2001]{tsc01}
{\sc {The International SNP Map Working Group}}, 2001 A map of human genome
  sequence variation containing 1.42 million single nucleotide polymorphisms.
\newblock Nature {\bf 409}: 928--933.

\end{thebibliography}

    \begin{figure}
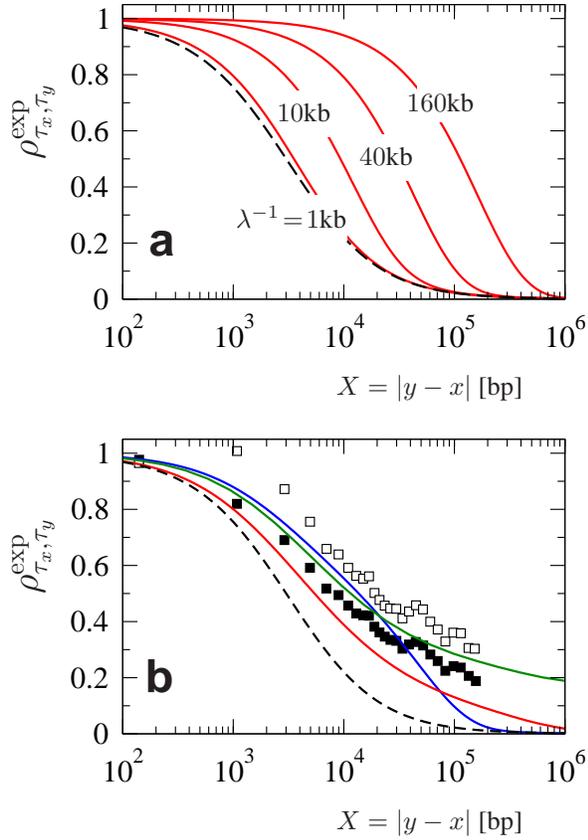

    \psfrag{c}{\Large\sfb a }
    \psfrag{y}{\large $\rho^{\rm exp}_{\tau_x,\tau_y}$}
    \psfrag{x}{$X = |y-x|$ [bp]}
    \psfrag{t1}{\hspace*{-1cm}$\lambda^{-1}\!=\!1$kb}
    \psfrag{t2}{\hspace*{-1.5mm}$10$kb}
    \psfrag{t3}{\hspace*{-2mm}$40$kb}
    \psfrag{t4}{\hspace*{-2mm}$160$kb}
    \includegraphics[width=7.5cm,clip]{fig2a.eps}
    \psfrag{d}{\Large\sfb b}
    \psfrag{x}{$X = |y-x|$ [bp]}
    \mbox{}\\[0.5cm]
    \includegraphics[width=7.5cm,clip]{fig2b.eps}

    \caption{\label{fig:2}
     \baselineskip=8mm
      Decorrelation of gene histories $\rho^{\rm exp}_{\tau_x,\tau_y}$ .
      {\bf a}
      Model I (red lines),
      $R=4Nr$, $N=10^4$, and $r=1.2$cM/Mb.
      The dashed line corresponds to constant recombination rate.
      {\bf b} Model II (blue line),
      $p=0.55,\lambda^{-1}=50$kb, $r=1.2$cM/Mb,
      and $L\gg X$, model III (red line), no fitting
      parameters, sex-averaged $p_X(C)$, model IV (green line) 
      for $\gamma=0.3$ (asymmetric split),
      empirical data (taken from Fig.~6a in \citep{rei02}, upper and
      lower confidence limits, points), constant recombination rate
      (dashed line).}
    \end{figure}

    \end{document}